\documentclass{PoS}

\title{Monte Carlo simulation of abelian gauge-Higgs lattice models using dual representations}

\ShortTitle{Gauge-Higgs lattice models with dual representations}

\author{\speaker{Alexander Schmidt},
Ydalia Delgado Mercado, Christof Gattringer%
         \thanks{Y.D.M and A.S. are members of the doctoral training program FWF DK 1203 ''{\sl Hadrons in Vacuum, Nuclei and Stars}''. Y.D.M.  is furthermore supported by the Research Executive Agency of the European Union 
under Grant Agreement number PITN-GA-2009-238353 (ITN STRONGnet). This work is partly supported also by DFG SFB TRR55.}\\
         \\
         Institut f\"ur Physik,
        Karl-Franzens-Universit\"at, 8010 Graz, Austria \\ \\
        \email{alexander.schmidt@uni-graz.at} \\ \email{ydalia.delgado-mercado@uni-graz.at} \\ \email{christof.gattringer@uni-graz.at}}

\abstract{\vspace{15mm}We study abelian gauge-Higgs models on the lattice and consider gauge groups Z$_3$ and U(1). For both cases the partition sums are mapped exactly to a dual representation where the degrees of freedom are surfaces for the gauge fields and loops of flux that may serve as boundaries for the surfaces represent the matter fields. Also at finite chemical potential the dual partition sums have only real and positive contributions and the complex action problem of the conventional representation is overcome in the dual approach. We apply a local Metropolis update for the dual degrees of freedom, as well as a generalization of the worm algorithm to bounded surfaces. Results that illustrate condensation phenomena as a function of chemical potential are discussed.}

\FullConference{The 30th International Symposium on Lattice Field Theory\\
                 June 24 - 29,  2012\\
                 Cairns, Australia}

\begin{document}

\section{Introductory remarks}
\noindent
Several interesting lattice field theories have a so-called complex action problem, i.e., in some parameter range the action $S$ has an imaginary part and the Boltzmann factor 
$\exp(-S)$ cannot be used as a weight factor in a Monte Carlo simulation. Examples are lattice QCD at finite density or lattice QCD with a theta term. In a quite remarkable series of developments  for several simpler systems the complex action problem was overcome in recent years by mapping them onto new degrees of freedom (dual representation), where the partition sum is a sum over real and positive contributions. The dual variables so far mainly consisted of loops that are subject to constraints. Such systems can be updated efficiently with algorithms based on the Prokof'ev Svistunov worm algorithm \cite{worm}.

In a recent paper \cite{graznew} we were able to solve the complex action problem for a new class of systems: Abelian gauge-Higgs models. In this case the dual degrees of freedom are surfaces for the gauge fields which are either open or bounded by flux lines representing the matter fields. In this contribution we report on these recent results and present ideas for further development of the concepts for Monte Carlo simulations with  dual representations.

\section{Abelian gauge-Higgs models as theories of surfaces}

\noindent
We here consider two abelian gauge-Higgs models, one based on the gauge group Z$_3$, the other one on U(1) with two flavors. The conventional action of the Z$_3$ gauge-Higgs model is given by 
\begin{equation} 
  S \; = \; - \frac{\beta}{2} \sum_x \sum_{\nu < \rho} 
  \Big[ U_{x,\nu\rho} + U_{x,\nu\rho}^\star \Big]
  - \eta \sum_{x,\nu} \! \left[ e^{\mu \delta_{\nu,4}} \phi_x^\star \, U_{x,\nu} \,\phi_{x+\widehat{\nu}}   +
    e^{- \mu \delta_{\nu,4}}  \phi_x^\star \, U_{x-\widehat{\nu}, \nu}^\star \, \phi_{x-\widehat{\nu}}  \right] \! \!  .
  \label{gaugeactionz3}
\end{equation}
The first part is the Wilson plaquette action where $U_{x,\nu\rho} = U_{x,\nu} U_{x + \widehat{\nu},\rho} U_{x + \widehat{\rho},\nu}^\star 
U_{x ,\rho}^\star$ is the plaquette for the link variables  $U_{x,\nu} \in $ Z$_3=\{e^{i2\pi/3},e^{-i2\pi/3},1\}$.
The second part consists of hopping terms for the Higgs field variables $\phi_x \in $ Z$_3$ made gauge invariant with link variables.
We work on lattices of size $N_s^3 \times N_t$ and use periodic boundary conditions for all directions. $\beta$ is the inverse gauge coupling and the factor
$\eta$ controls the coupling between the Higgs- and the gauge fields.
Essentially it plays the role of an inverse Higgs mass: If the Higgs field is
infinitely heavy ($\eta = 0$) it  decouples from the dynamics of the system.
The hopping terms in the 4-direction (= temporal direction) 
are coupled to the chemical potential $\mu$. In the conventional representation 
the model then has a complex action problem at $\mu > 0$.  
The partition sum is obtained by summing the Boltzmann factor over all possible
field configurations
$ Z  = \sum_{\{\phi,U\}} \exp(-S)$.

\begin{figure}[b!]
  \begin{center}
    \includegraphics[height=2.5cm]{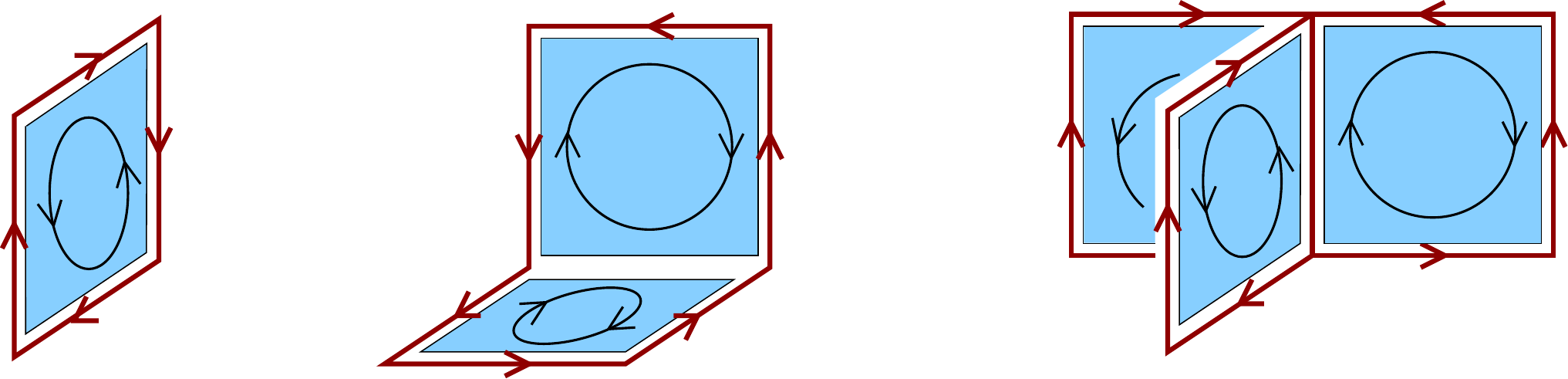}
    \caption{Examples of admissible configurations in the dual representation of the Z$_3$ model.}
  \end{center}
  \label{admissiblez3}
\end{figure}

The complex action problem can be overcome completely \cite{graznew} by
 mapping the partition sum to a sum over new degrees of freedom: Plaquette occupation numbers $p_{x,\nu\rho}
\in \{-1,0,1\}$ which are attached to the plaquettes of the lattice and represent the gauge degrees of freedom  
(see also \cite{others} for related results in pure abelian gauge theory), and fluxes 
$k_{x,\nu} \in \{-1,0,1\}$ assigned to the links of the lattice which represent the matter field. The partition sum can be rewritten exactly into the form
\begin{equation}
Z \; = \; C \sum_{\{p\}}  \sum_{\{k\}} {\cal C}_P[p,k] \,  {\cal C}_F[k] \,  {\cal W}_P[p] \,  {\cal W}_F[k] \; .
\label{ZfinalZ3}
\end{equation}
Here $C$ is a constant depending on the parameters $\beta$ and $\mu$, and the sum runs over all configuration of the $p$ and $k$ variables. Each configuration is weighted with real and positive factors
\begin{equation}
 {\cal W}_P[p] \;  = \;  \prod_{x,\nu < \rho} B_\beta^{\; \, p_{x,\nu\rho}} \quad , \quad
 {\cal W}_F[k]  \;  = \; \prod_x \left(M_{k_{x,4}} \;  \prod_{j=1}^3  B_\eta^{\; |k_{x,j}|}  \right)\;,
\end{equation}
where
\begin{eqnarray}
B_\beta &  = &  \frac{e^{\beta}\ -\ e^{-\beta/2}}{e^{\beta}\ +\ 2e^{-\beta/2}} \quad , \quad
  B_\eta \; = \; \frac{e^{2\eta} - e^{-\eta}}{e^{2\eta} + 2 e^{-\eta} }\; , \\
  M_k \; &=& \; \frac{1}{3} \Big[ e^{2\eta \cosh(\mu)} + \; 2 e^{-\eta \cosh(\mu) } \cos\Big( \sqrt{3} \eta \sinh(\mu) - k \frac{2\pi}{3}\Big) \Big] \; , \; k = -1,0,+1 \;.
  \nonumber
\end{eqnarray}
In addition the configurations are subject to the constraints
\begin{eqnarray}
{\cal C}_F[k] & = &  \prod_x  T \Big( \sum_\nu \big[ k_{x,\nu} - k_{x-\widehat{\nu},\nu} \big]   \Big)   \; ,  \nonumber \\
{\cal C}_P[p,k]  & = & \prod_{x,\nu}  T \left( \; \sum_{\rho:\nu < \rho}\! \big[ p_{x,\nu\rho} - p_{x-\widehat{\rho},\nu\rho}\big]
- \sum_{\rho:\rho<\nu} \! \big[ p_{x,\rho\nu} - p_{x-\widehat{\rho},\rho\nu}\big]
  + k_{x,\nu} \right) \; .
\label{constraintsz3}  
\end{eqnarray}  
In the two constraints $T(n)$ denotes the triality function which equals 1 if $n$ is a multiple of 3 and vanishes otherwise. The first constraint is a product
over all sites and at each site forces the total matter flux to vanish modulo 3. The second constraint is a product over all links and forces the combined plaquette and matter flux at each link 
to vanish modulo 3. The corresponding admissible configurations have the interpretation of closed surfaces of plaquette variabes or surfaces bounded by matter flux, and in 
Fig.~\ref{admissiblez3}  we show simple examples of admissible configurations.

The other gauge-Higgs system we consider is based on the gauge group U(1). In the conventional representation the fields are the link variables $U_{x,\nu} \in$ U(1) and the two flavors ($f = 1,2$) of complex valued scalar fields $\phi_x^{(f)}$ with opposite charges. The action reads,
\begin{eqnarray}
  S & = & -  \beta \sum_{x} \sum_{\nu < \rho}  \, \mbox{Re} \; U_{x,\nu} \, U_{x + \widehat{\nu}, \rho} \, U_{x + \widehat{\rho}, \nu}^\star \, U_{x,\rho}^\star  \; + \;
\sum_{f=1}^2 \sum_{x}\! \Bigg[ \kappa^{(f)} \! \mid \!\! {\phi^{(f)}_{x}} \!\! \mid^2
\;  + \; \lambda^{(f)} \! \mid \!\! {\phi^{(f)}_{x}} \!\! \mid^4  \Bigg]  \\
& & \hspace{30mm} -  \;  \sum_{x,\nu}\! \Bigg[ e^{\, \mu^{(1)} \delta_{\nu, 4}} \, {\phi^{(1)}_{x}}^\star \, U_{x,\nu} \,  {\phi^{(1)}_{x+\widehat{\nu}}}   
\;   + \; e^{-\mu^{(1)} \delta_{\nu, 4} } \, {\phi^{(1)}_{x}}^\star \, U_{x - \widehat{\nu},\nu}^{\;\star} \,  {\phi^{(1)}_{x-\widehat{\nu}}} \Bigg] 
  \nonumber \\
& & \hspace{30mm} -  \;  \sum_{x,\nu}\! \Bigg[ e^{\, \mu^{(2)} \delta_{\nu, 4}} \, {\phi^{(2)}_{x}}^\star \, U_{x,\nu}^{\; \star} \,  {\phi^{(2)}_{x+\widehat{\nu}}}   
\;   + \; e^{-\mu^{(2)} \delta_{\nu, 4} } \, {\phi^{(2)}_{x}}^\star \, U_{x - \widehat{\nu},\nu} \,  {\phi^{(2)}_{x-\widehat{\nu}}} \Bigg] \; .
  \nonumber
\end{eqnarray}
Again we use the Wilson-plaquette action for the gauge fields. For the Higgs fields the parameters are quartic couplings $\lambda^{(f)}$ and 
$\kappa^{(f)} = 8 + {m^{(f)}}^2$, with $m^{(f)}$ the bare masses.  The hopping-terms in 4-direction are coupled to chemical potentials 
$\mu^{(f)}$ and again we have a complex action problem when $\mu^{(f)} \neq 0$.
 
Similar to the Z$_3$-model it is possible to rewrite the partition sum exactly in terms of a dual representation with 
plaquette variables $p$ for the gauge fields and two sets of flux variables $k^{(f)}$ and $l^{(f)}$ for the two flavors ($f = 1,2$) of matter fields,
\begin{eqnarray}
  \hspace{-0.5cm} Z \; = \hspace{-0.5cm} \sum_{\{p, k^{(1)}, l^{(1)}, k^{(2)}, l^{(2)}\}} \hspace{-0.5cm} {\cal W}_P[p] \, {\cal W}_F^{(1)}[k^{(1)},l^{(1)}] \, {\cal W}_F^{(2)}[k^{(2)},l^{(2)}] \, {\cal C}_P[p,k^{(1)},k^{(2)}] \, {\cal C}_F[k^{(1)}] \, {\cal C}_F[k^{(2)}] \; , \\
k^{(1)},\;k^{(2)} \in (-\infty,\infty),\quad  l^{(1)},\;l^{(2)} \in [0,\infty), \quad p \in (-\infty,\infty) \; . \nonumber
\end{eqnarray}
The additional sets of flux variables $ l^{(1)},\;l^{(2)} $ appear because for the U(1) case the matter fields also have a radial degree of freedom which is absent in the Z$_3$ case. However, it is important to note that the link variables $l^{(f)}$ for the radial degrees of freedom do not enter in the constraints.  As for the Z$_3$ case it turns out that in the dual form of the U(1) model all weights  ${\cal W}_P,  {\cal W}_F^{(1)},  {\cal W}_F^{(2)}$ (which we do not display explicitly here) are real and positive and the complex action problem is solved. The constraints ${\cal C}_F$ for the flux variables $k^{(f)}$ are the same as in 
Eq.~(\ref{constraintsz3}) for the Z$_3$ case, with the only difference that the triality function $T(n)$ is replaced by a Kronecker delta $\delta_{n,0}$ which here we denote as $\delta(n)$. The constraint  ${\cal C}_P$ that links the plaquette and flux variables reads
\begin{equation}
{\cal C}_P[p,k^{(1)},k^{(2)}]  \; = \; \prod_{x,\nu}  \delta \, \left( \,\sum_{\rho: \nu < \rho}\! \big[ p_{x,\nu\rho} - p_{x-\widehat{\rho},\nu\rho}\big]
- \!\sum_{\rho:\rho<\nu} \! \big[ p_{x,\rho\nu} - p_{x-\widehat{\rho},\rho\nu}\big]
  + \, k^{(1)}_{x,\nu}  \, -  \, k^{(2)}_{x,\nu} \right) ,
\label{constraintsu1}  
\end{equation}
where the different signs for  $k^{(1)}$ and $k^{(2)}$ reflect the opposite charges of the two flavors. 

In this study we use thermodynamical observables that
can be obtained as derivatives of $\ln Z$ with respect to the various couplings, such as the plaquette $\langle U \rangle$, the particle number density $n$, and the corresponding susceptibilities,
\begin{equation}
  \langle U \rangle =  \frac{1}{6 N_s^3 N_t} \, \frac{\partial  \, \ln Z}{\partial \beta} \; ,
  \; \chi_U  = \frac{1}{6 N_s^3 N_t} \, \frac{\partial^2  \, \ln Z}{\partial \beta^2}  \; , \; 
  n \; = \; \frac{1}{N_s^3 N_t}  \frac{\partial  \, \ln Z}{\partial \mu} \; , \;
  \chi_{n} \; = \; \frac{1}{N_s^3 N_t} \, \frac{\partial^2  \, \ln Z}{\partial \mu^2} \; .
\end{equation}
The derivatives can easily be worked out also in the dual representations and the observables are then obtained as combinations of moments of the dual variables. 
 
\begin{figure}[b!]
  \begin{center}
    \includegraphics[height=3cm]{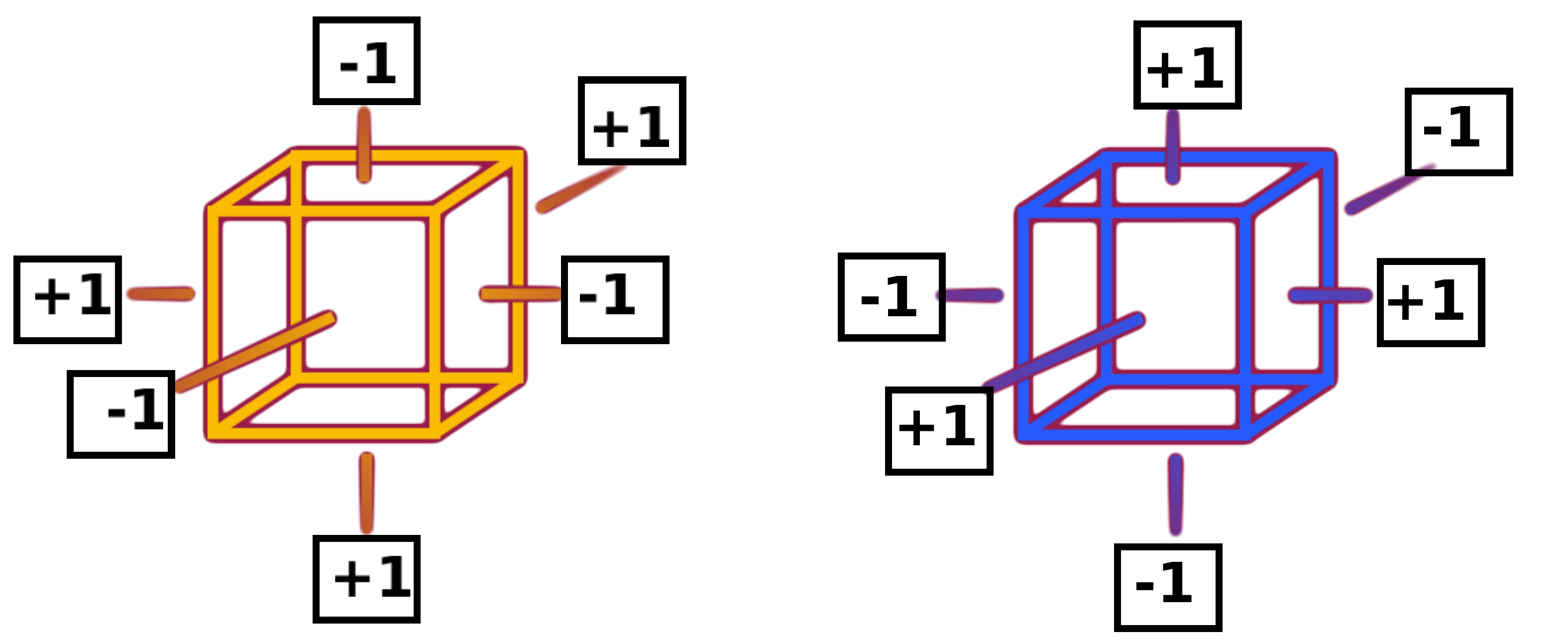} \hspace{0.3cm}
    \includegraphics[height=2cm]{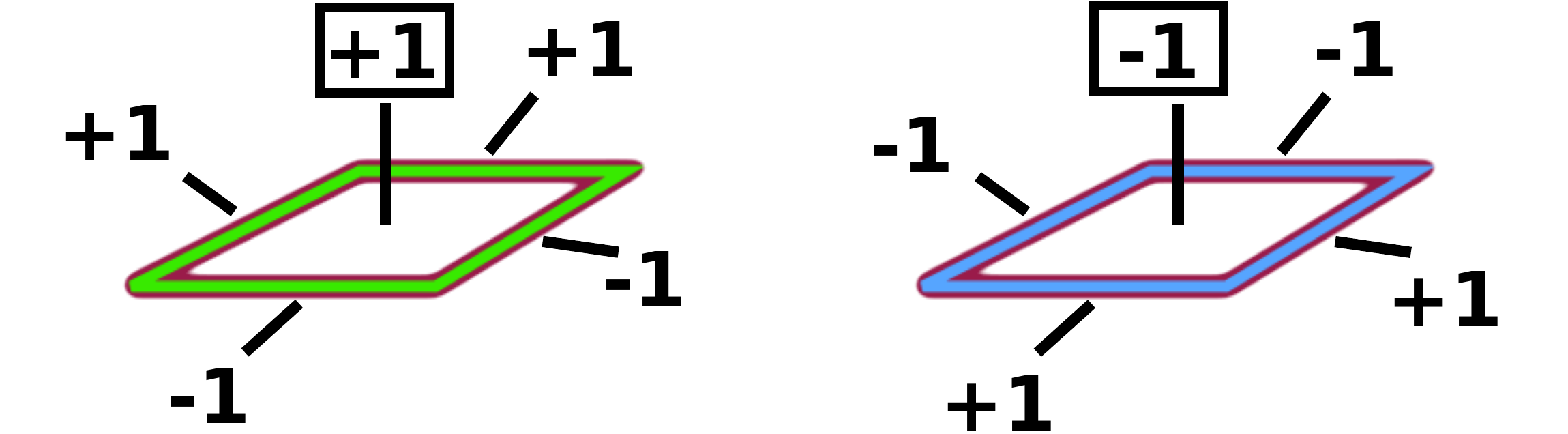}
   \caption{Cube updates (two lhs. plots) and plaquette updates (rhs.) used in our local Metropolis algorithm.}
       \label{localMC}
  \end{center}
\end{figure}

\section{Update strategies for the dual variables}

\noindent
In order to do simulations in the dual representation we need a Monte-Carlo algorithm which  generates admissible configurations of the dual variables, i.e., configurations that obey all constraints. The simplest approach is a local Metropolis update where plaquette and flux variables on cubes and plaquettes are changed according to a simple pattern, such that the constraints remain intact. We use two updates which we refer to as cube and plaquette updates. The cube update only changes the plaquette variables $p$ on the faces of a 3-cube embedded in four dimension according to one of the two patterns in the two lhs.~plots of Fig.~\ref{localMC}, where for the case of Z$_3$ all additions are understood modulo 3.    It is easy to see that such a change leaves the constraints ${\cal C}_P$ intact, and since no flux variables are changed also the constraints 
${\cal C}_F$.  A full sweep of cube updates consists of offering one of the two cube updates of  Fig.~\ref{localMC} with equal probability to all 3-cubes of the lattice and accepting the changes with the usual Metropolis probability. To update the flux variables, in the plaquette update we offer a
joint update of one plaquette variable and the fluxes on the links of that plaquette according to the possibilities shown in the two rhs.~plots of Fig.~\ref{localMC} (again addition is modulo 3 for Z$_3$). Also here one easily shows that the constraints  ${\cal C}_P$  and ${\cal C}_F$ remain intact, and a full sweep of plaquette updates consists of offering the change to all plaquettes of the lattice. For the case of the 2-flavor U(1) model there exist two different plaquette sweeps for the two flavors, and the unconstrained variables $l^{(f)}$ are updated with conventional Monte Carlo sweeps.

\begin{figure}[t]
\begin{center}
\hspace*{-4mm}
\includegraphics[width=13.5cm,clip]{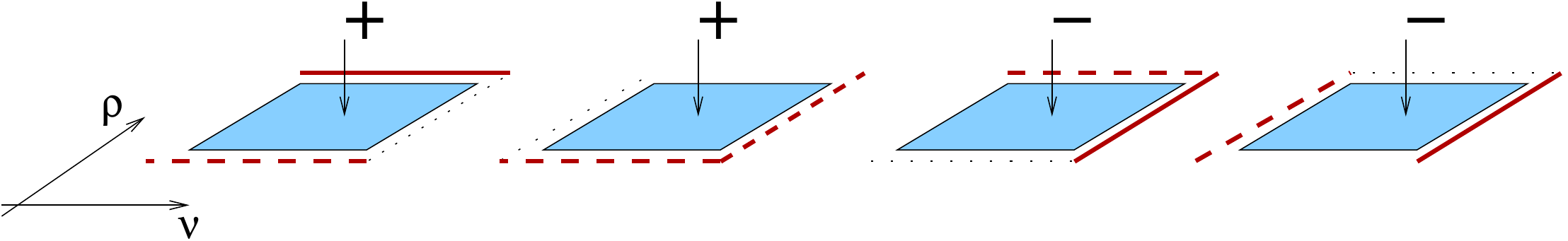}
\end{center}
\vspace{-4mm}
\caption{Examples of  segments used in the surface worm algorithm.
The plaquette occupation numbers are changed as indicated by the signs. 
The links marked with full (dashed) lines are changed by $+1$ ($-1$). The empty link shows
where the segment is attached and the dotted link is the new position of the defect where the constraints are violated.}
\label{segments}
\end{figure}

A second, more efficient strategy is a generalization of the worm algorithm \cite{worm} to surfaces with boundaries \cite{grazworm}. After changing the flux at a randomly chosen link, which creates a defect in the constraints, one adds so-called segments which consist of a plaquette and two of its links. The corresponding plaquette number and the fluxes on the two links are changed (see Fig.~\ref{segments} for examples of segments). Attaching such a segment to the defect propagates the defect on the lattice until the worm decides to heal the violated constraints at the defect by adding the missing flux.
It may be shown \cite{grazworm}, that the generalized surface worm algorithm clearly outperforms the local updates for a wide range of parameters.   

\section{Selected results}

\begin{figure}[t]
  \centering
  \includegraphics[width=64mm,clip]{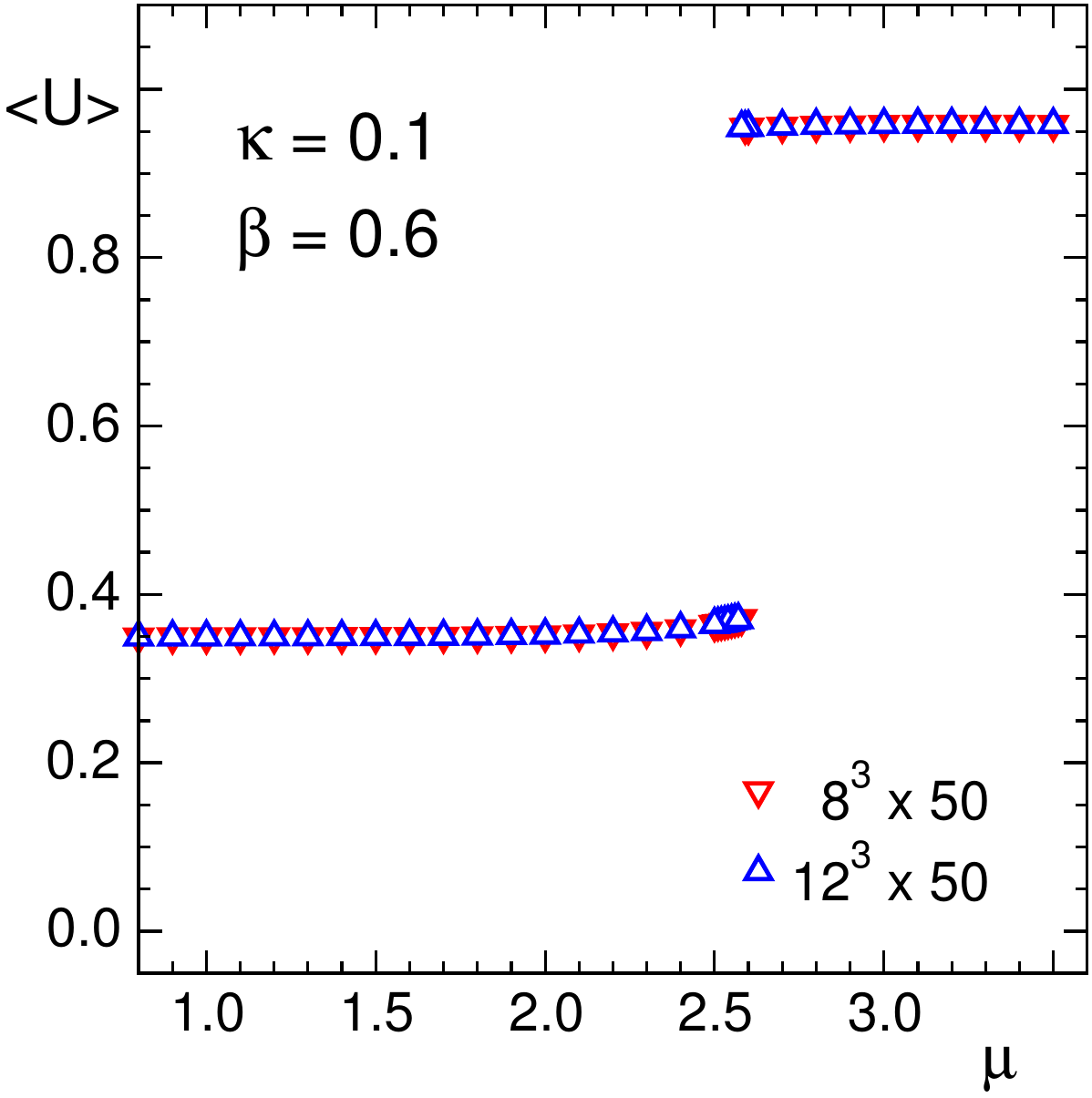}
  \includegraphics[width=85mm,clip]{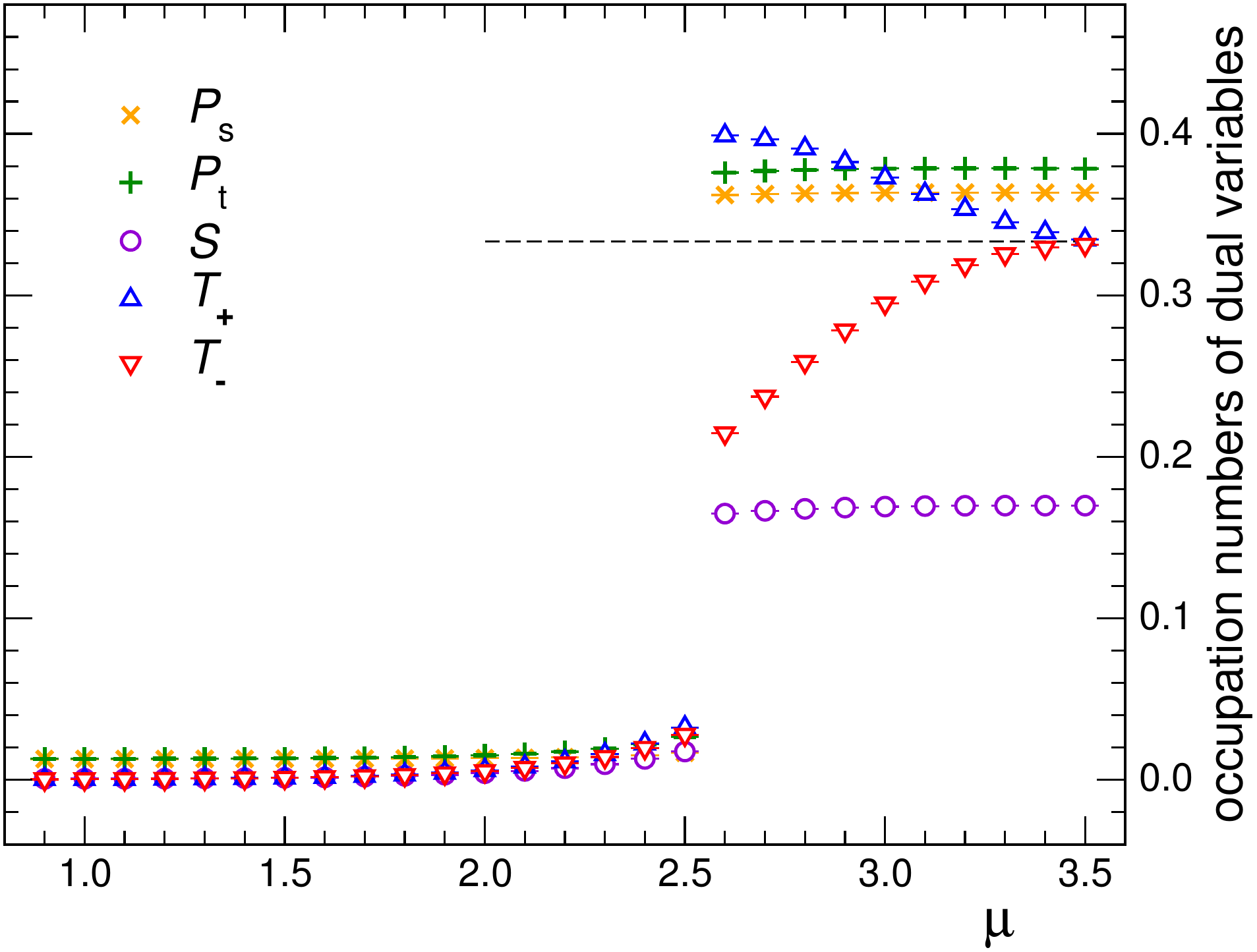}
  \caption{Lhs.: Results for $ \langle U \rangle$ 
    in the strong coupling region of the
    Z$_3$ gauge-Higgs model ($\eta = 0.1$ and
    $\beta = 0.6$) versus the chemical potential $\mu$. Rhs.: The corresponding occupation numbers for temporal and spatial plaquettes ($P_t, P_s$),  for spatial fluxes ($S$), and for positive and negative temporal fluxes ($T_+, T_-$) . 
    \label{condensationz3}}
\end{figure}

\noindent 
The results for the Z$_3$ gauge-Higgs model were presented in detail in our recent paper \cite{graznew}. From a physics point of view the interesting aspect of the analysis of the dual model is the possibility to study a theory with gauge interactions at weak coupling for low temperature and high density. In such a situation one expects condensation phenomena that give rise to phase transitions as a function of the chemical potential (see \cite{condensation} for examples of related condensation phenomena at strong coupling or with non-gauge interactions). 
In Fig.~\ref{condensationz3} we provide an example for such a condensation: The lhs.\ plot shows the plaquette expectation value $\langle U \rangle$ as a function of $\mu$, which obviously undergoes a pronounced first order transition near $\mu_c \sim 2.6$. The rhs.~plot shows the corresponding average plaquette and flux occupation numbers, illustrating that the transition can be understood as the condensation of fluxes
and plaquette occupation numbers.    

\begin{figure}[b]
  \centering
  \includegraphics[width=1.0\textwidth,clip]{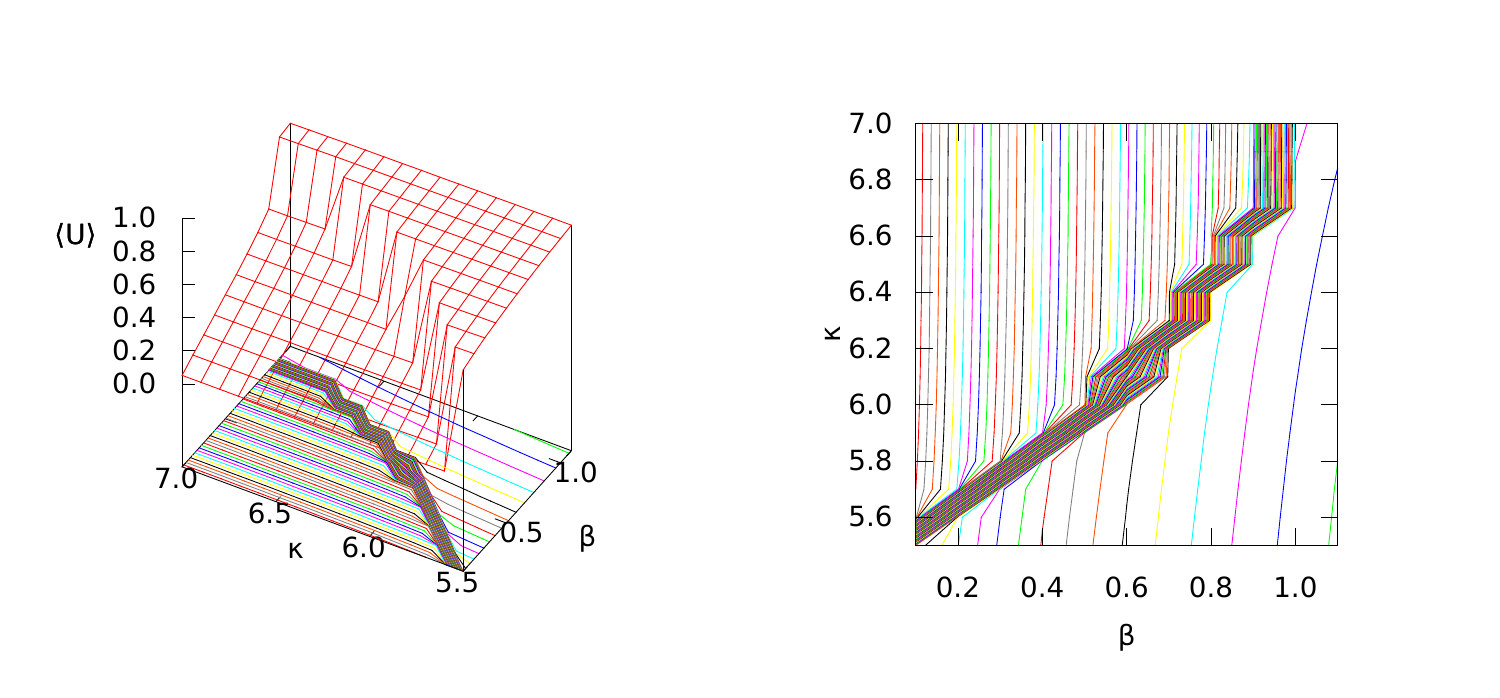}
  \caption{In the lhs.~plot we show a 3-D plot of the plaquette expectation value $\langle U \rangle$ as function of $\kappa^{(1)} = \kappa^{(2)} = \kappa$ and $\beta$ for $\lambda^{(f)}=0$
  and $\mu^{(f)} = 0$. 
    The contour lines in the base of the lhs.~plot and on the rhs.~are equally spaced lines of constant $\langle U \rangle$.}
  \label{coolphase}
\end{figure}

Even more interesting is the physics of the U(1) gauge-Higgs model, which we currently study for $\lambda^{(1)} = \lambda^{(2)} = 0$ and degenerate masses $m^{(1)} = m^{(2)}$, i.e., $\kappa^{(1)} = \kappa^{(2)} = \kappa$. At $\kappa = \infty$, which corresponds to pure U(1) gauge theory, the system undergoes a first order transition at $\beta_c \sim 1$, which is signalled by a pronounced disccontinuity in $\langle U \rangle$. This transition persists also for the two flavor model at finite $\kappa$ and in Fig.~\ref{coolphase}  we show our results from the dual simulation for  $\langle U \rangle$ 
as a function of $\beta$ and $\kappa$ at $\mu^{(1)} = \mu^{(2)} = 0$. One clearly observes the transition that separates the strong coupling phase
from the weak coupling phase with $\langle U \rangle \sim 1$

\begin{figure}[t]
  \centering
  \includegraphics[width=1\textwidth,clip]{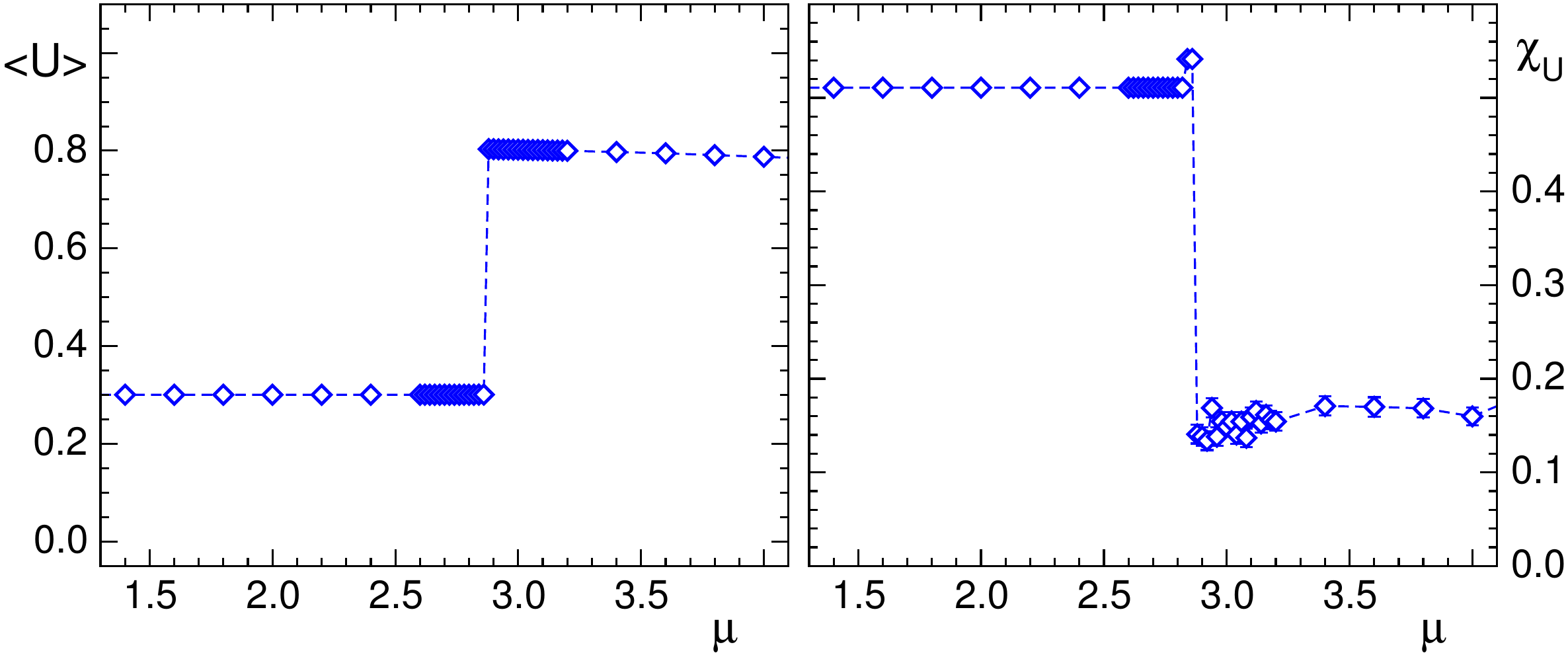}
  \caption{Plaquette expectation value $\langle U \rangle$ (lhs.) and plaquette suscpetibility $\chi_U$ (rhs.) as function of the chemical
    potential ($8^3 \times 100$, $\kappa=6.4$, $\beta=0.6$, $\lambda=0$).}
  \label{silvertrans}
\end{figure}

In Fig.~\ref{silvertrans} we plot the plaquette expectation value $\langle U \rangle$ and the corresponding susceptibility $\chi_U$ as functions
of the chemical potential $\mu$ for the parameter set $\kappa=6.4$, $\beta=0.6$ (strong coupling regime). We find that the observables are independent 
of the chemical potential up to  $\mu_c \sim 2.9$ (Silver Blaze phenomenon), where the system undergoes a  pronounced first order transition, which 
as in the Z$_3$ case corresponds to a condensation of the dual degrees of freedom.

\end{document}